\documentclass{article}

\everydisplay{\mathsurround 0pt} 

\usepackage[a4paper]{geometry}
\usepackage{amsmath,amssymb,amsthm,mathtools}
\usepackage[colorlinks,citecolor=blue,urlcolor=blue]{hyperref}
\usepackage[affil-it]{authblk}
\usepackage[round]{natbib}
\usepackage{setspace}
\usepackage[utf8]{inputenc}

\usepackage{pdflscape}
\usepackage{relsize}
\usepackage{amsmath}
\usepackage{comment}
\usepackage{eucal} 
\usepackage{latexsym} 
\usepackage{amssymb} 
\usepackage{accents}
\usepackage{enumitem}
\usepackage{cancel}
\usepackage{rotating}
\usepackage{url}
\usepackage{bigints}
\usepackage{graphicx}
\usepackage{subfig}

\usepackage{lscape}

\usepackage{ae,lmodern}  
\usepackage[french, english]{babel}
\usepackage[utf8]{inputenc}
\usepackage[T1]{fontenc}
\usepackage{booktabs}

\usepackage{multirow}

\usepackage{float}

\newcommand{\Xb}{\ensuremath{\mathbf{X}}}
\newcommand{\Yb}{\ensuremath{\mathbf{Y}}}  
\newcommand{\xb}{\ensuremath{\mathbf{x}}}

\newcommand{\thetab}{\ensuremath{\pmb{\theta}}}

\newcommand{\Sigmab}{\ensuremath{\pmb{\Sigma}}}

\newcommand{\Deltab}{\ensuremath{\pmb{\Delta}}}

\newcommand{\Rb}{\ensuremath{\mathbf{R}}}

\newcommand{\Sb}{\ensuremath{\mathbf{S}}}
\newcommand{\Ib}{\ensuremath{\mathbf{I}}}

\newcommand{\Rd}{\ensuremath{\mathbb{R}^d}} 
\newcommand{\Ik}{\ensuremath{\mathbf{I}_d}}

\newcommand{\Ub}{\ensuremath{\mathbf{U}}}

\newcommand{\Sigmanb}{\ensuremath{\hat{\pmb{\Sigma}}}}

\newcommand{\Gammab}{\ensuremath{\pmb{\Gamma}}} 

\newcommand{\R}{\mathbb{R}}

\newcommand*{\Scale}[2][4]{\scalebox{#1}{$#2$}}%

\usepackage{xcolor}
\usepackage{verbatim}

\newcommand{\proglang}[1]{\textsf{#1}}
\newcommand{\pkg}[1]{{\fontseries{b}\selectfont #1}}
\newcommand{\code}[1]{\textsf{#1}}

\usepackage{thumbpdf,lmodern}

\usepackage{framed}

\usepackage{authblk}

\title{Elliptical Symmetry Tests in \proglang{R}} 

\author[1,2]{Sla\dj ana Babi\'c}
\affil[1]{Ghent University, Department of Applied Mathematics, Computer Science and Statistics, 9000 Ghent, Belgium, \{sladana.babic, christophe.ley, marko.palangetic\}@ugent.be}
\affil[2]{Vlerick Business School, B-1210 Brussels, Belgium}
\author[1]{Christophe Ley
}

\author[1]{Marko Palangeti\'c}

\date{}

\begin{document}
\maketitle

\begin{abstract}
\indent 
 The assumption of elliptical symmetry has an important role in many theoretical developments and applications, hence it is of primary importance to be able to test whether that assumption actually holds true or not.  Various tests have been proposed in the literature for this problem. To the best of our knowledge, none of them has been implemented in \textbf{R}.
 The focus of this paper is the implementation of several well-known tests for elliptical symmetry together with some recent tests. 
 We demonstrate the testing procedures with a real data example.

\end{abstract}


{\it MSC2010 subject classification}: 62-04

\section{Introduction}
Let $\Xb_1,\ldots,\Xb_n$ denote a sample of $n$ i.i.d.~$d$-dimensional observations. A $d$-dimensional random vector $\Xb$ is said to be elliptically symmetric about some location parameter $\thetab\in\R^d$ if its density~$\underline{f}$ is of the form 
\begin{equation}\label{def1}
\xb\mapsto{\underline{f}(\xb;\thetab,\Sigmab,f)}=c_{d,f}|\Sigmab|^{-1/2}f\left(\|\Sigmab^{-1/2}(\xb-\thetab)\|\right),\qquad\mathbf{x}\in\mathbb{R}^d,
\end{equation}
where $\Sigmab\in\mathcal{S}_d$ (the class of symmetric positive definite real $d\times d$ matrices) is a {\it scatter} parameter, $f:\mathbb{R}^+_0\to\mathbb{R}^+$ is an a.e.\  strictly positive function called {\it radial density}, and $c_{d,f}$ is a normalizing constant depending on~$f$ and the dimension~$d$. Many well-known and widely used multivariate distributions are elliptical. The multivariate normal, multivariate Student $t$, multivariate power-exponential, symmetric multivariate stable, symmetric multivariate Laplace, multivariate logistic, multivariate Cauchy, and multivariate symmetric general hyperbolic distribution are all examples of elliptical distributions. The family of elliptical distributions has several appealing properties. For instance, it has a simple stochastic representation, clear parameter interpretation, it is closed under affine transformations, and its marginal and conditional distributions are also elliptically symmetric: see \cite{paindaveine2014elliptical} for details. Thanks to its mathematical tractability and nice properties, it became a fundamental assumption in multivariate analysis and many applications. Numerous statistical procedures therefore rest on the assumption of elliptical symmetry: one- and $K$-sample location and shape problems   \citep{UmRandles98, hallin2002optimal, hallin2006semiparametrically1, hallin2006semiparametrically2}, 
serial dependence and  time series \citep{HPd2004AoS}, 
one- and $K$-sample  principal component problems   \citep{hallin2010optimal, HPdV2014JASA},   multivariate tail estimation \citep{dominicy2017multivariate}, to cite but a few. Elliptical densities  are also considered in portfolio theory \citep{OR83}, capital asset pricing models \citep{hodgson2002testing}, semiparametric density estimation \citep{liebscher2005semiparametric}, graphical models \citep{vogel2011elliptical},  and many other areas. 

\noindent
Given the omnipresence of the  assumption of elliptical symmetry, it is essential to be able to test whether  that assumption actually holds true or not for the data at hand. Numerous tests have been proposed in the literature, including \citet{beran1979testing}, \citet{baringhaus1991testing},   \citet{koltchinskii2000testing},  \citet{manzotti2002statistic}, \citet{schott2002testing},  \citet{huffer2007test}, \citet{Cassart07} and \citet{BGHL19}. Tests for elliptical symmetry based on Monte Carlo simulations can be found in \citet{diks1999test} and \citet{zhu2000nonparametric};  \citet{li1997some} recur to graphical methods and \citet{zhu2004} build conditional tests. We refer  the reader to  \citet{serfling2004multivariate} and \citet{sakhanenko2008testing} for extensive reviews and  performance comparisons. To the best of our knowledge, none of these tests is available in the open software \textbf{R}. The focus of this paper is to close this gap by implementing  several well-known tests for elliptical symmetry together with some recent tests. 
The test of \cite{beran1979testing} is
neither distribution-free nor affine-invariant; moreover, there are no practical guidelines to the choice of the basis functions involved in the test
statistic. Therefore, we opt not to include it in the package. \citet{baringhaus1991testing} proposes a Cram\'er-von Mises type test for spherical symmetry based on the independence between norm and direction. \citet{dyckerhoff2015depth} have shown by simulations that this test can be used as a test for elliptical symmetry in dimension 2. This test assumes the location parameter to be known and its asymptotic distribution is not simple to use (plus no proven validity in dimensions higher than 2), hence we decided not to include it in the package. Thus, the tests suggested by \citet{koltchinskii2000testing}, \citet{manzotti2002statistic}, \citet{schott2002testing}, \citet{huffer2007test}, \citet{Cassart07} and \citet{BGHL19} are implemented in the package \pkg{ellipticalsymmetry}. 

\noindent
 This paper describes the tests for elliptical symmetry that have been implemented in the \pkg{ellipticalsymmetry} package, together with a detailed description of the functions that are available in the package. The use of the implemented functions is illustrated using financial data.

 
 \section{Testing for elliptical symmetry}
In this section, we focus on tests for elliptical symmetry that have been implemented in our new \pkg{ellipticalsymmetry} package. Besides formal definitions of test statistics and limiting distributions, we also explain the details on computing.

\subsection{Test by Koltchinskii and Sakhanenko}
 \cite{koltchinskii2000testing} develop a class of omnibus bootstrap tests for unspecified location that are affine invariant and consistent against any fixed alternative. The estimators of the unknown parameters are as follows:
 $\hat{\thetab} = n^{-1}\sum_{i=1}^{n}\Xb_i$ and $\hat{\Sigmab} = n^{-1}\sum_{i=1}^{n}(\Xb_i - \hat{\thetab})(\Xb_i - \hat{\thetab})'$. Define $\Yb_i = \Sigmanb^{-1/2}(\Xb_i-\hat{\thetab})$ and let~$\mathcal{F}_B$ be a class of Borel functions from $\Rd$ to $\R$. Their test statistics are functionals (for example, sup-norms) of the stochastic process\vspace{-1mm}
$$
n^{-1/2}\sum_{i=1}^{n}\left(f(\Yb_i) - m_{f}(||\Yb_i||)\right), \vspace{-1mm}
$$
where $f \in \mathcal{F}_B$ and $m_f(\rho)$ is the average value of $f$ on the sphere with radius~$\rho>0$. Several examples of classes $\mathcal{F}_B$ and test statistics based on the sup-norm of the above process are considered in~\cite{koltchinskii2000testing}. Here, we restrict our attention to~$\mathcal{F}_B:=\left\{I_{0<||\xb||\leq t}\psi\left(\frac{\xb}{||\xb||}\right):\psi\in G_l,||\psi||_2\leq1,t>0\right\}$ where $I_{A}$ stands for the indicator function of   $A$,  $G_l$ for the linear space of spherical harmonics of degree less than or equal to $l$ in $\R^d$, and~$||\cdot||_2$ is the~$L^2$-norm on    the unit sphere $\mathcal{S}^{d-1}$ in $\R^d$. With these quantities in hand, the test statistic becomes
$$
Q_{{KS}}^{(n)} := n^{-1/2}\max_{1\leq j \leq n}\left(\sum_{s=1}^{\text{dim}(G_l)}\left(\sum_{k=1}^{j}\psi_s\left(\frac{\Yb_{[k]}}{||\Yb_{[k]}||}\right) - \delta_{s1}\right)^2 \right)^{1/2},
$$
where $\Yb_{[i]}$ denotes the $i$th order statistic from the sample {$\Yb_1, \dots, \Yb_n$ ordered according to their $L^2$-norm}, $\{\psi_s, \ s =1, \dots,\, \text{dim}(G_l)\}$ denotes an orthonormal basis of $G_l,\ \psi_1 = 1$, and $\delta_{ij} = 1$ for $i=j$ and $0$ otherwise. The test statistic is relatively simple to construct if we have formulas for spherical harmonics. 
In dimension $2$ spherical harmonics coincide with sines and cosines on the unit circle. The detailed construction of the test statistic $Q_{{KS}}^{(n)}$ for dimensions $2$ and $3$ can be found in
\citet{sakhanenko2008testing}. In order to be able to use $Q_{{KS}}^{(n)}$  in higher dimensions, we need corresponding formulas for spherical harmonics. Using recursive formulas from \citet{muller1966lecture} and equations given in \citet{manzotti2001spherical} we obtained spherical harmonics of degree one to four in arbitrary dimension. The reader should bare in mind that larger degree leads to better power performance of this test. A drawback of this test is that it requires bootstrap procedures to obtain critical values.

\noindent
In our \textbf{R} package, this test can be run using a function called \verb+KoltchinskiiSakhanenko()+. The syntax for this function  is very simple:
\begin{verbatim}
    KoltchinskiiSakhanenko(X, R=1000, nJobs = -1),
\end{verbatim}
where \code{X} is an input to this function consisting of a data set which must be a matrix and \code{R} stands for the number of bootstrap replicates. The default number of replicates is set to $1000$. The \code{nJobs} argument represents the number of CPU cores to use for the calculation. This is a purely technical option which is used to speed up the computation of bootstrap based tests. The default value $-1$ indicates that all cores except one are used. 

\subsection{The MPQ test}
\citet{manzotti2002statistic} develop a test based on spherical harmonics. The estimators of the unknown parameters are the sample mean denoted as $\hat{\thetab}$ and the unbiased sample covariance matrix given by $\hat{\Sigmab} = \frac{1}{n-1}\sum_{i=1}^{n}(\Xb_i - \hat{\thetab})(\Xb_i - \hat{\thetab})'$. Define again $\Yb_i = \Sigmanb^{-1/2}(\Xb_i-\hat{\thetab})$.
 When the $\Xb_{i}'s$ are elliptically symmetric, then $\Yb_i/||\Yb_i||$ should be  uniformly distributed on the unit sphere, and \citet{manzotti2002statistic} chose this property as the basis of their test. The uniformity of the standardized vectors $\Yb_i/||\Yb_i||$ can be checked in different ways. \cite{manzotti2002statistic} opt to verify this uniformity using spherical harmonics.
 For a fixed $\varepsilon > 0$, let $\rho_n$ be the $\varepsilon$ sample quantile of $||\Yb_1||, \dots, ||\Yb_n||$.  Then, the test statistic is
$$
Q_{{MPQ}}^{(n)} = n \mathop{\Scale[2]{\sum}}_{h \in \mathcal{S}_{jl}}\bigg( \frac{1}{n} \mathop{\Scale[2]{\sum}}_{i=1}^n h\bigg(\frac{\Yb_i}{||\Yb_i||}\bigg)I(||\Yb_i|| > \rho_{n})\bigg)^2
$$
for $l\geq j \geq 3$, where $\mathcal{S}_{jl} = \bigcup_{j\leq i\leq l} \mathcal{H}_i$
and $\mathcal{H}_i$ is the set of spherical harmonics of degree $i$. 
 In the implementation of this test we used spherical harmonics of degree $3$ and $4$. 
 The asymptotic distribution of the test statistic $Q_{{MPQ}}^{(n)}$ is $(1 - \varepsilon) \chi$, where $\chi$ is a variable with a chi-squared distribution with $\nu_{jl}$ degrees of freedom, where $\nu_{jl}$ denotes the total number of functions in $\mathcal{S}_{jl}$.  
 Note that $Q_{{MPQ}}^{(n)}$ is only a necessary condition statistic for the null hypothesis of elliptical symmetry and therefore this test does not have asymptotic power  against all alternatives. 
In the \pkg{ellipticalsymmetry} package, this test is implemented as the \verb+MPQ+ function with the following syntax
\begin{verbatim}
    MPQ(X, epsilon = 0.05)
\end{verbatim}
As before, \code{X} is a numeric matrix that represents the data while \code{epsilon} is an option that allows the user to indicate the proportion of points $\Yb_i$ close to the origin which will not be used in the calculation. By doing this, extra assumptions on the radial density in \eqref{def1} are avoided. 
The default value of \code{epsilon} is set to 0.05.

\subsection{Schott's test}

\citet{schott2002testing} develops a Wald-type test for elliptical symmetry based on the analysis of covariance matrices. The test compares the sample fourth moments with the expected theoretical ones under ellipticity.
Given that the test statistic involves consistent estimates of the covariance matrix of the sample fourth moments, the existence of eight-order moments is required.  Furthermore, the test has very low power against several alternatives. The final test statistic is of a simple form, even though it requires lengthy notations.
 
 
 
 \noindent
 For an elliptical distribution with mean $\thetab$ and covariance matrix $\Sigmab$, the fourth moment defined as ${\pmb M_4} = E\{(\Xb- \thetab)(\Xb- \thetab)'\otimes (\Xb- \thetab)(\Xb- \thetab)'\}$, with $\otimes$ the Kronecker product, has the form
\begin{align}
\label{eq:M4}
    {\pmb M_4} = (1 + \kappa)((\Ib_{d^2} + K_{dd})(\Sigmab \otimes \Sigmab) + vec(\Sigmab)vec(\Sigmab)')
\end{align}
where $K_{dd}$ is a commutation matrix \citep{magnus1988linear}, $\Ib_d$  is the $d\times d$ identity matrix, and $\kappa$ is a scalar which can be expressed using the characteristic function of the elliptical distribution. Here the $vec$ operator stacks all components of a $d\times d$ matrix $\pmb M$ on top of each other to yield the $d^2$ vector ${\rm vec}(\pmb M)$.
Let $\hat{\Sigmab}$ denote the usual unbiased sample covariance matrix and $\hat{\thetab}$ the sample mean.
A simple estimator of ${\pmb M_4}$ is given by 
 $
  \hat{\pmb M}_4 = \frac{1}{n}\sum_{i=1}^n(\Xb_i - \hat{\thetab})(\Xb_i - \hat{\thetab})'\otimes (\Xb_i - \hat{\thetab})(\Xb_i - \hat{\thetab})'
 $
 and its standardized version is given by
 $$
 { \hat{\pmb M}_{4*}}= (\hat{\Sigmab}^{-1/2'}\otimes \hat{\Sigmab}^{-1/2'}){\hat{\pmb M}_4} (\hat{\Sigmab}^{-1/2}\otimes \hat{\Sigmab}^{-1/2}).
 $$
Then, an estimator of $vec({\pmb M}_4)$ is constructed
as $  {\pmb G} = vec({\pmb N}_4)vec({\pmb N}_4)'vec(\hat{\pmb M}_{4*})/(3d(d+2))$, and it is consistent if and only if ${\pmb M}_4$ is of the form (\ref{eq:M4}); here ${\pmb N}_4$ represents the value of ${\pmb M}_4$ under the multivariate standard normal distribution.  Note that the asymptotic mean of $ \pmb{v} =  n^{1/2}( vec(\hat{\pmb M}_{4*}) - {\pmb G})$ is $0$ if and only if (\ref{eq:M4}) holds and this expression is used to construct the test statistic. Denote the estimate of the asymptotic covariance matrix of $n^{1/2}\pmb{v}$ as $\hat{\pmb \Phi}$. The Wald test statistic is then formalized as $ T = \pmb{v}' \hat{\pmb \Phi}^{-} \pmb{v}$, where $\hat{\pmb \Phi}^{-}$ is a generalized inverse of $\hat{\pmb \Phi}$. For more technical details we refer the reader to Section 2 in \citet{schott2002testing}.
In order to define Schott's test statistic, we further have to define the following quantities:
 $$
 (1 + \hat{\kappa}) = \frac{1}{nd(d+2)} \sum_{i=1}^n\{(\Xb_i - \hat{\thetab})'\hat{\Sigmab}^{-1}(\Xb_i - \hat{\thetab})\}^2
 $$
 
 $$
 (1 + \hat{\eta}) = \frac{1}{nd(d+2)(d+4)}\sum_{i=1}^n\{(\Xb_i - \hat{\thetab})'\hat{\Sigmab}^{-1}(\Xb_i - \hat{\thetab})\}^3
 $$
 
 $$
 (1 + \hat{\omega}) = \frac{1}{nd(d+2)(d+4)(d+6)}\sum_{i=1}^n\{(\Xb_i - \hat{\thetab})'\hat{\Sigmab}^{-1}(\Xb_i - \hat{\thetab})\}^4.
 $$
 
 \noindent
Moreover, let $\hat{\beta}_1 = (1 + \hat{\omega})^{-1}/24,\enspace \hat{\beta}_2 = -3a\{ 24 (1 + \hat{\omega})^2 +12(d+4)a(1 + \hat{\omega})\}^{-1} $, $a = (1 + \hat{\omega}) + (1 + \hat{\kappa})^3 - 2(1 + \hat{\kappa})(1 + \hat{\eta})$. Finally, the test statistic becomes
$$
T = n\left[ \hat{\beta}_1 \text{tr}(\hat{\pmb M}_{4*}^{2}) + \hat{\beta}_2\text{vec}(\Ib_d)' \hat{\pmb M}_{4*}^{2}\text{vec}(\Ib_d) - \{3\hat{\beta}_1 + (d+2)\hat{\beta}_2\}d(d+2)(1+\hat{\kappa})^2 \right].
$$
It has an asymptotic chi-squared distribution with degrees of freedom
$\displaystyle \nu_d = d^2 + \frac{d(d-1)(d^2 + 7d - 6)}{24} - 1.$

\noindent
The Schott test can be performed in our package by using the function \verb+Schott+ with the very simple syntax \verb+Schott(X)+, where \code{X} is a numeric matrix of data values.

\subsection{Test by Huffer and Park}
\citet{huffer2007test} propose a Pearson chi-square type test with multi-dimensional cells. Under the null hypothesis of ellipticity the cells have asymptotically equal expected cell counts and after determining the observed cell counts, the test statistic is easily computed. Let $\hat{\thetab}$  be the sample mean and $\hat{\Sigmab} = n^{-1}\sum_{i=1}^{n}(\Xb_i - \hat{\thetab})(\Xb_i - \hat{\thetab})'$ the sample covariance matrix. Define $\Yb_i = \Rb(\Xb_i - \hat{\thetab}),$
where the matrix $\Rb=\Rb(\hat{\Sigmab})$ is a function of $\hat{\Sigmab}$ such that $\Rb\hat{\Sigmab} \Rb = \Ib_d$. Typically $\Rb = \hat{\Sigmab}^{-1/2}$ as for the previous tests. However, Huffer and Park suggest to use the Gram-Schmidt transformation because that will lead to standardized data whose joint distribution does not depend on $\thetab$ or $\Sigmab$. In order to compute the test statistic, the space $\Rd$ should be divided into $c$ spherical shells centered at the origin such that each shell contains an equal number of the scaled residuals $\Yb_i$. 
The next step is to divide $\Rd$ into $g$ sectors such that for any pair of sectors there is an orthogonal transformation mapping one onto the other. Therefore, the $c$ shells and $g$ sectors divide $\Rd$ into $gc$ cells which, under elliptical symmetry, should contain $n/(gc)$ of the vectors $\Yb_i$. The test statistic then has the simple form
$$
HP_n = \sum_{\pi}(U_{\pi} - np)^2/(np),
$$
where $U_{\pi}$ are cell counts for $\pi = (i,j)$ with $1\leq i\leq g$ and $1\leq j \leq c$ and $p=1/(gc)$.

\noindent
In the \textbf{R} package we are considering three particular ways to partition the space: using (i) the $2^d$ orthants, (ii) permutation matrices and (iii) a cyclic group consisting of rotations by angles which are multiples of $2\pi/g$. The first two options can be used for any dimension, while the angles are supported only for dimension 2. 
Huffer and Park's test can be run using a function called \verb+HufferPark+. The syntax, including all options, for the function \verb+HufferPark+ is for instance
\begin{verbatim}
HufferPark(X, c, R = NA, sector = "orthants", g = NA, nJobs = -1).
\end{verbatim}
We will now provide a detailed description of its arguments.
\code{X} is an input to this function consisting of a data set.
\code{sector} is an option that allows the user to specify the type of sectors used to divide the space. Currently supported options are  \code{"orthants"}, \code{"permutations"} 
and \code{"bivariateangles"}, the last one being available only in dimension $2$. The \code{g} argument indicates the number of sectors. 
The user has to choose \code{g} only if \code{sector = "bivariateangles"} and it denotes the number of regions used to divide the plane. In this case, regions consist of points whose angle in polar coordinates is between $2(m-1)\pi/g$ and $2m\pi/g$ for $m \in \{1 \dots g\}$. 
If \code{sector} is set to \code{"orthants"}, then \code{g} is fixed and equal to $2^d$, while for \code{sector = "permutations"} \code{g} is $d!$. 
No matter what type of sectors is chosen, the user has to specify the number of spherical shells that are used to divide the space, which is \code{c}.
 The value of \code{c} should be such that the average cell counts $\displaystyle n/(gc)$ are not too small. Several settings with different sample size,
 and different values of $g$ and $c$ can be found in the simulation studies presented in Sections 4 and 5 of \citet{huffer2007test}. As before, \code{nJobs} represents the number of CPU cores to use for the calculation. The default value \code{-1} indicates that all cores except one are used.
 
 \noindent
The asymptotic distribution is available only under \code{sector = "orthants"} when the underlying distribution is close to normal. It is a linear combination of chi-squared random variables and it depends on eigenvalues of congruent sectors used to divide the space $\Rd$. 
Otherwise, bootstrap procedures are required and the user can freely choose the number of bootstrap replicates, denoted as \code{R}. 
Note that by default \code{sector} is set to \code{"orthants"} and \code{R = NA}. 
\subsection{Pseudo-Gaussian test}
\citet{Cassart07} and \cite{CHPd2008JSPI} construct Pseudo-Gaussian tests for specified and unspecified location  that are most efficient against  a multivariate form of Fechner-type  asymmetry (defined in \citet{Cassart07}, Chapter 3). 
These tests are based on Le Cam's asymptotic theory of statistical experiments.
We start by describing the specified-location Pseudo-Gaussian test. The unknown parameter $\Sigmab$ is estimated by using \citet{tyler1987distribution}'s estimator of scatter which we simply denote by $\hat{\Sigmab}$. 
Let $m_{k}(\thetab,\Sigmab) := n^{-1}\sum_{i=1}^{n}(\|\Sigmab^{-1/2}(\Xb_i-\thetab)\|)^k$, $U_i(\thetab,\Sigmab) := \frac{\Sigmab^{-1/2}(\Xb_i-\thetab)}{\|\Sigmab^{-1/2}(\Xb_i-\thetab)\|}$ and \vspace{-1mm}
\begin{align*}\nonumber\Sb^{\Ub}_i(\thetab,\Sigmab) := ((\Ub_{i1}(\thetab,\Sigmab))^2 \text{sign}&(\Ub_{i1}(\thetab,\Sigmab)), \dots, (\Ub_{id}(\thetab,\Sigmab))^2 \text{sign}(\Ub_{id}(\thetab,\Sigmab)))^\prime . \vspace{-1mm}
\end{align*}
The test statistic then has the simple form
\vspace{-1mm}
$$Q_{{p{\cal G},\thetab}}^{(n)} = \frac{d(d+2)}{3n{m_4}(\thetab,\hat\Sigmab)}\sum_{i,j=1}^n  (\|\hat\Sigmab^{-1/2}(\Xb_i-\thetab)\|)^2 (\|\hat\Sigmab^{-1/2}(\Xb_j-\thetab)\|)^2\Sb'^{\Ub}_i(\thetab,\hat\Sigmab)\Sb^{\Ub}_j(\thetab,\hat\Sigmab)
\vspace{-1mm}$$ 
and follows asymptotically a chi-squared  distribution $\chi^2_d$ with $d$ degrees of freedom. Finite moments of order four are required.



\noindent
In most cases the assumption of a specified center is however unrealistic. \citet{Cassart07}  therefore proposes also a test for the scenario when location is not specified. 
The estimator of the unknown $\thetab$ is the sample mean denoted by $\hat{\thetab}$. Let $\Yb_i = \hat\Sigmab^{-1/2} (\Xb_i - \hat\thetab)$. The test statistic takes on the guise
$$
Q_{{p{\cal G}}}^{(n)} := (\Deltab_{{{\cal G}}}(\hat\thetab,\hat\Sigmab))'(\Gammab_{{{\cal G}}}(\hat\thetab,\hat\Sigmab))^{-1}\Deltab_{{{\cal G}}}(\hat\thetab,\hat\Sigmab),
$$ where
$$
\Deltab_{{{\cal G}}}(\hat\thetab,\hat\Sigmab)=  n^{-1/2}\sum_{i=1}^{n}\|\Yb_i\|\left(c_d(d+1)m_{1}(\hat\thetab,\hat\Sigmab)\Ub_i(\hat\thetab,\hat\Sigmab)-\|\Yb_i\|\Sb^{\Ub}_i(\hat\thetab,\hat\Sigmab)\right)
$$
and
\begin{align*}
  &\Gammab_{{{\cal G}}}(\hat\thetab,\hat\Sigmab) :=\\ &\left(\frac{3}{d(d+2)}m_{4}(\hat\thetab,\hat\Sigmab) - 2c_{d}^2(d+1)m_{1}(\hat\thetab,\hat\Sigmab)m_{3}(\hat\thetab,\hat\Sigmab) + c_{d}^{2}\frac{(d+1)^2}{d}(m_{1}(\hat\thetab,\hat\Sigmab))^2m_{2}(\hat\thetab,\hat\Sigmab)\right)\Ik  
\end{align*}
with $c_d = 4\Gamma(d/2)/((d^2-1)\sqrt{\pi}\Gamma(\frac{d-1}{2}))$, $\Gamma(\cdot)$ being the Gamma function.
 The test rejects the null hypothesis of elliptical symmetry at asymptotic level $\alpha$ whenever the test statistic 
$
Q_{{p{\cal G}}}^{(n)} 
$ 
exceeds $\chi^2_{d;1-\alpha}$, the upper $\alpha$-quantile of a $\chi^2_d$ distribution. We refer   to Chapter 3 of \cite{Cassart07} for formal details.

\noindent
This test can be run in our package by calling the function  \verb+pseudoGaussian+ with the  simple syntax 
\begin{verbatim}
   pseudoGaussian(X, location = NA).
\end{verbatim}
Besides \code{X} which is a numeric matrix of data values, now we have an extra argument \code{location} which allows the user to specify the known location. The default is set to \code{NA} which means that the unspecified location test will be performed unless the user specifies location.

\subsection{SkewOptimal test}
Recently, \citet{BGHL19} proposed a new test for elliptical symmetry both for specified and unspecified location. These tests are based on Le Cam's asymptotic theory of statistical experiments and are optimal against generalized skew-elliptical alternatives (defined in Section 2 of said paper) but they remain quite powerful under a much broader class of non-elliptical distributions. 

\noindent
The test statistic for the specified location scenario has a very simple form and an asymptotic chi-squared distribution. The test rejects the null hypothesis whenever
$
Q_{\thetab}^{(n)}=n (\bar{\Xb}-\thetab)'\hat\Sigmab^{-1}(\bar{\Xb}-\thetab)
$
exceeds   the $\alpha$-upper quantile $\chi^2_{d;1-\alpha}$. Here, $\hat{\Sigmab}$ is  \citet{tyler1987distribution}'s estimator of scatter and $\bar{\pmb X}$ is the sample mean. 

\noindent
When the location is not specified, \citet{BGHL19} propose tests that have a simple asymptotic chi-squared distribution under the null hypothesis of ellipticity,  are affine-invariant, computationally fast, have a simple and intuitive form,
only require finite moments of order 2, and offer much flexibility in the choice of the radial density $f$ at which optimality (in the maximin sense) is achieved. Note that the Gaussian $f$ is excluded, though, due to a singular information matrix; see \citet{BGHL19}. We implemented in our package the test statistic  based on the radial density $f$ of the multivariate $t$ distribution, multivariate power-exponential and multivariate logistic, though in principle any non-Gaussian choice for $f$ is possible. The test requires lengthy notations, but its implementation is straightforward. For the sake of generality, we will derive the test statistic for a general (but fixed) $f$, and later on provide the expressions of $f$ for the three special cases implemented in our package.  Let $\varphi_f(x) = -\frac{f'(x)}{f(x)}$ and $\Yb_i = \hat\Sigmab^{-1/2} (\Xb_i - \hat\thetab)$ where $\hat\thetab$ is the sample mean. In order to construct the test statistic, we first have to define the  quantities  \begin{align*}
   \displaystyle {\Deltab}_{f}(\hat\thetab,\hat\Sigmab) =  2n^{-1/2}\dot{\Pi}(0)\sum_{i=1}^n\left[\|\Yb_i\|-\frac{d}{ \widehat{\mathcal{K}}_{d,f}(\hat\thetab,\hat\Sigmab)}\varphi_f(\|\Yb_i\|)\right]\frac{\Yb_i}{\|\Yb_i\|}
   \end{align*} and  
 $$\widehat{\Gammab}_{f}(\hat\thetab,\hat\Sigmab):=\frac{{4(\dot{\Pi}(0))^2}}{nd}\sum_{i=1}^n\left[\|\Yb_i\|-\frac{d}{ \widehat{\mathcal{K}}_{d,f}(\hat\thetab,\hat\Sigmab)}\varphi_f(\|\Yb_i\|)\right]^2\Ik$$
   where
  $
          \widehat{\mathcal{K}}_{d,f}(\hat\thetab,\hat\Sigmab) := \frac{1}{n}\sum_{i=1}^n \left[ \displaystyle \varphi_f'(\|\Yb_i\|) + \frac{d-1}{\|\Yb_i\|}\varphi_f(\|\Yb_i\|) \right]
      $ and $\Pi$ is the cdf of the standard normal distribution (we use $\dot{\Pi}(\cdot)$ for the derivative).
Finally, the test statistic is of the form
$  Q^{(n)}_f:= ({\Deltab}_{f}(\hat\thetab,\hat\Sigmab))' (\widehat{\Gammab}_{f}(\hat\thetab,\hat\Sigmab))^{-1}{\Deltab}_{f}(\hat\thetab,\hat\Sigmab)$ and it has a chi-squared distribution with~$d$ degrees of freedom. The test is valid under the entire semiparametric hypothesis of elliptical symmetry with unspecified center and uniformly optimal against any type of generalized skew-$f$ alternative. 

\noindent
From this general expression, one can readily derive the test statistics for specific choices of $f$. In our case, the radial density of the multivariate Student $t$ distribution corresponds to       $f(x) = (1 + \frac{1}{\nu}x^2)^{-(\nu + d)^2}$, where $\nu \in (0,\infty)$ represents the degrees of freedom, while that of the multivariate logistic distribution is given by $\displaystyle f(x)= \frac{\exp{(-x^2)}}{[1+\exp{(-x^2})]^2}$ and of the multivariate power-exponential by $\displaystyle f(x)= \exp{\left(-\frac{1}{2}x^{2\beta}\right)}$, where $\beta \in (0,\infty)$ is a parameter related to kurtosis. 

\noindent
These tests can be run in \textbf{R} using a function called \verb+SkewOptimal+ with the  syntax
\begin{verbatim}
   SkewOptimal(X, location = NA, f = "t", param = NA)
\end{verbatim}

\noindent
Depending on the type of the test some of the input arguments are not required. \code{X} and \code{location} are the only input arguments for the specified location test, and have the same role as for the Pseudo-Gaussian test. As before, the default value for \code{location} is set to \code{NA} which implies that the unspecified location test will be performed unless the user specifies location. For the unspecified location test, besides the data matrix \code{X}, the input arguments are \code{f} and \code{param}. The \code{f} argument is a string that specifies the type of the radial density based on which the test is built. Currently supported options are \code{"t"}, \code{"logistic"} and \code{"powerExp"}. Note that the default is set to \code{"t"}. The role of the \code{param} argument is as follows. If \code{f = "t"} then \code{param} denotes the degrees of freedom of the multivariate $t$ distribution. Given that the default radial density is \code{"t"}, it follows that the default value of \code{param} represents the degrees of freedom of the multivariate $t$ distribution and it is set to $4$. Note also that the degrees of freedom have to be greater than $2$.
If \code{f = "powerExp"} then \code{param} denotes the kurtosis parameter $\beta$, in which case the value of \code{param} has to be different from $1$ because $\beta = 1$ corresponds to the multivariate normal distribution. The default value is set to $0.5$.

\subsection{Time complexity}
We conclude the description of tests for elliptical symmetry by comparing their time complexity  in terms of the big O notation \citep{cormen2009introduction}. More concretely, we are comparing the number of simple operations that are required to evaluate the test statistics and the p-values. Table~\ref{time} summarizes the time complexity of the implemented tests. 

\noindent
The test of Koltchinskii and Sakhanenko is computationally more demanding than the bootstrap version of the test of Huffer and Park. Among unspecified location tests that do not require bootstrap procedures, the most computationally expensive test  is the MPQ test under the realistic assumption that $n > d$. Regarding the specified location tests we can conclude that the Pseudo-Gaussian test is more computationally demanding than the SkewOptimal test. Note that both the test of Koltchinskii and Sakhanenko and the MPQ test are based on spherical harmonics up to degree $4$. In case we would use spherical harmonics of higher degrees, the tests would  of course become even more computationally demanding. 

\noindent
We have seen that several tests require bootstrap procedures and therefore are by default computationally demanding. Such tests require the calculation of the statistic on the resampled data $R$ times in order to get the p-value, where $R$ is the number of bootstrap replicates. Consequently, the time required to obtain the p-value in such cases is $R$ times the time to calculate the test statistic. For the tests that do not involve bootstrap procedures, the p-value is calculated using the inverse of the cdf of the asymptotic distribution under the null hypothesis, which is considered as one simple operation. The exception here is the test of Huffer and Park whose asymptotic distribution is more complicated and includes $O(c)$ operations where $c$ is an integer and represents an input parameter for this test.

\begin{table}[H]
\centering
\caption{Time complexity of the various tests for elliptical symmetry}
\begin{tabular}{lll}
\toprule
                   & statistics &  p-value \\ 
\midrule
KoltchinskiiSakhanenko       &$O(n \log n + nd^5)$&$O(Rn \log n + Rnd^5)$\\ 
MPQ                &$O(n \log n + nd^5)$&    $O(1)$  \\
Schott             & $O(nd^2 + d^6)$ &     $O(1)$    \\ 
HufferPark       & $O(nd^2 + d^3)$  &  $O(c)$         \\ 
HufferPark (bootstrap)     & $O(nd^2 + d^3)$  &$O(Rnd^2 + Rd^3)$\\ 
PseudoGaussian (specified location) &$O(n^2d + nd^2 + d^3)$& $O(1)$    \\ 
PseudoGaussian & $O(nd^2 + d^3)$ &    $O(1)$     \\ 
SkewOptimal (specified location)    & $O(nd + d^3)$ &    $O(1)$     \\ 
SkewOptimal    & $O(nd^2 + d^3)$ &    $O(1)$     \\ 
\bottomrule
\end{tabular}
\label{time}
\end{table}

\section{Illustrations using financial data}
Mean-Variance analysis was introduced by \cite{markowitz} as a model for portfolio selection. In this model, the portfolio risk expressed through the historical volatility is minimized for a given expected return, or the expected return is maximized given the risk. The model is widely used for making portfolio decisions, primarily because it can be easily optimized using quadratic programming techniques. However, the model has some shortcomings among which the very important one  that it does not consider the prior wealth of the investor that makes decisions. This prior wealth is important since it influences the satisfaction that an investor has from gains. For example, the gain of 50\$ will not bring the same satisfaction to someone whose wealth is 1\$ as to someone whose wealth is 1000\$. This satisfaction further affects the decision-making process in portfolio selection. Because of that and other financial reasons, a more general concept of expected utility maximization is used (see e.g. \cite{schoemaker2013experiments}). However, the expected utility maximization is not an easy optimization problem, and some additional assumptions must be made in order to solve it. Hence, despite the expected utility maximization being more general, the mean-variance approach is still used due to its computational simplicity. \cite{chamberlain1983characterization} showed that the two approaches coincide if the returns are elliptically distributed. In other words, under elliptical symmetry the mean-variance optimization solves the expected utility maximization for any increasing concave utility function. 
Therefore, we want to test if the assumption of elliptical symmetry holds or not for financial return data. The data set that we analyze contains daily stock log returns of $3$ major equity market indexes from North America: S\&P 500 (US), TSX (Canada) and IPC (Mexico). The sample consists of 5369 observations, from January 2000 through July 2020. To remove temporal dependencies by filtering, following the suggestion of \cite{lombardi2009indirect}, GARCH(1,1) time series models were fitted to each series of log-returns.

\begin{figure}[H]
  \begin{center}  \caption{North America indexes (S\&P, TSX and IPC)}\label{rollingwindow}
  \vspace{0.5cm}
\subfloat[SkewOptimal]{\includegraphics[angle=0,width=0.5\linewidth]{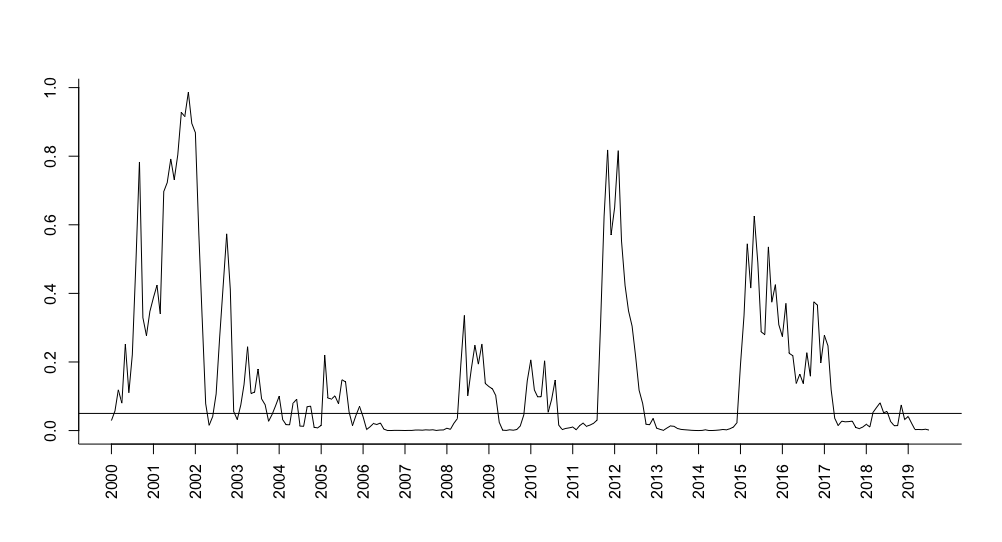}}
\subfloat[Pseudo-Gaussian]{\includegraphics[angle=0,width=0.5\linewidth]{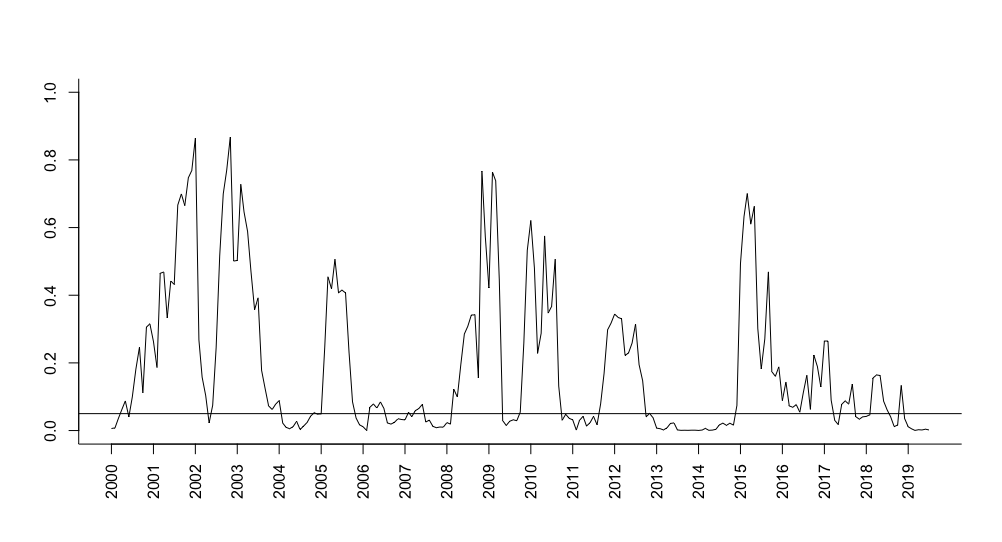}}\\
\subfloat[KoltchinskiiSakhanenko]{\includegraphics[angle=0,width=0.5\linewidth]{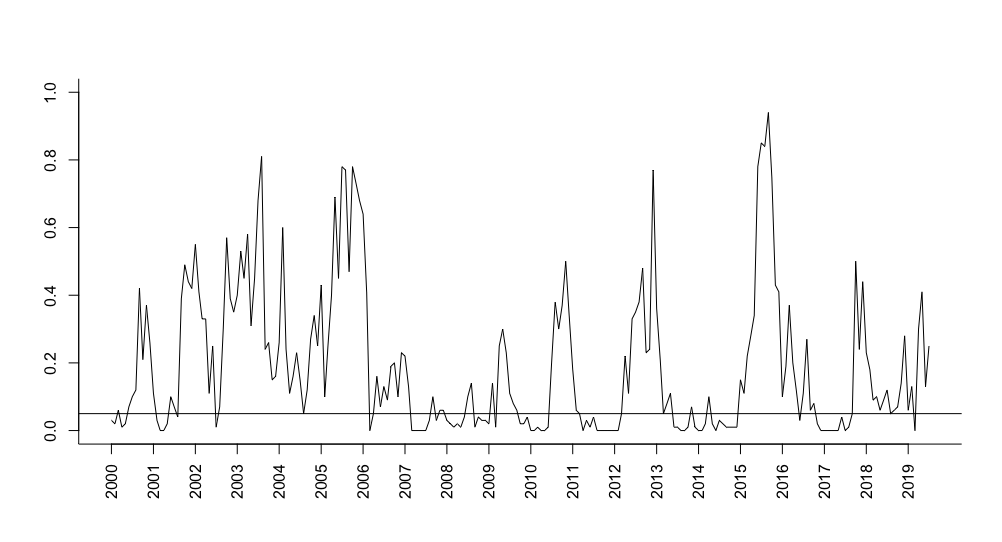}}
\subfloat[MPQ]{\includegraphics[angle=0,width=0.5\linewidth]{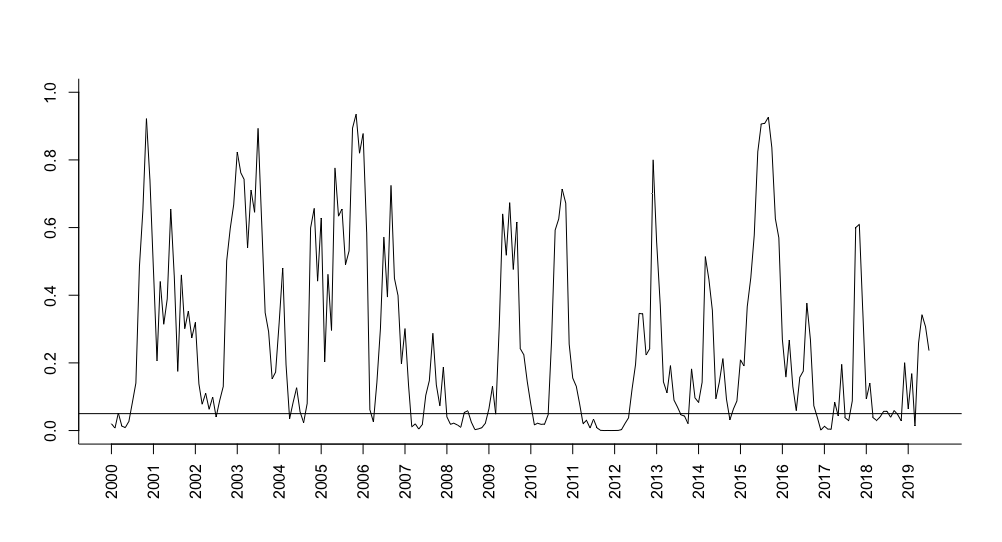}}\\
\subfloat[HufferPark]{\includegraphics[angle=0,width=0.5\linewidth]{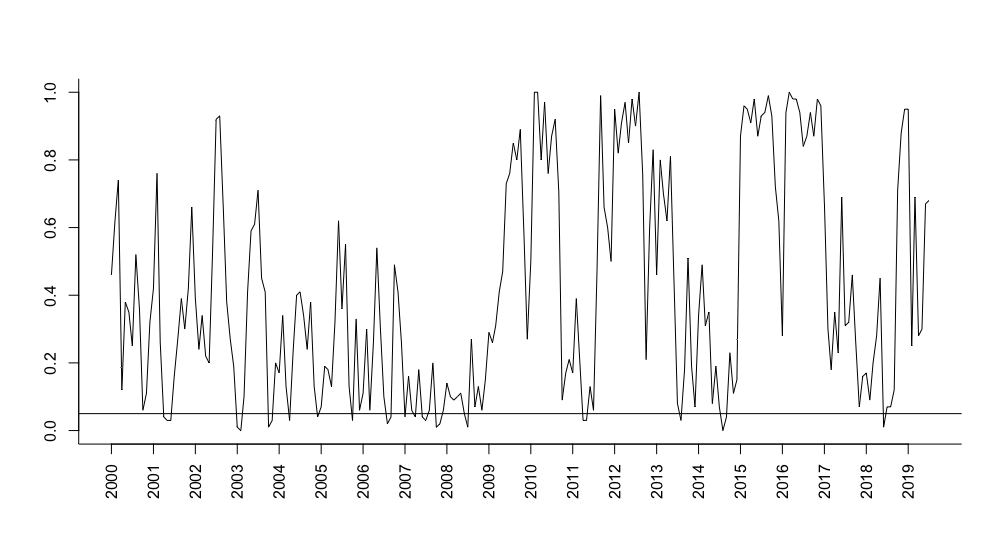}} 
\subfloat[Schott]{\includegraphics[angle=0,width=0.5\linewidth]{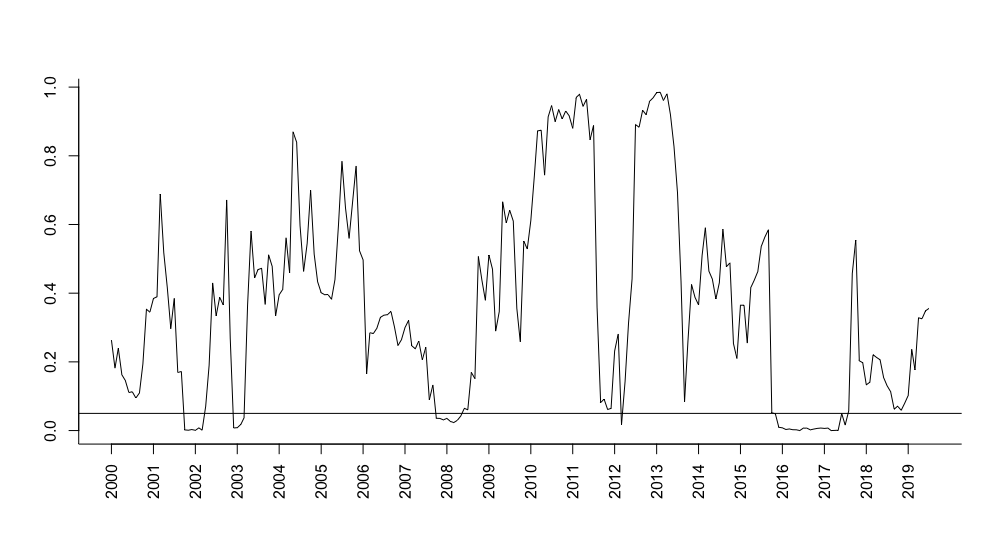}} \\
\begin{tabular}{p{12.5cm}}
\scriptsize The plots show the p-values of the corresponding tests for all rolling windows that we considered between 2000 and 2020. The years on the x-axis mark the rolling windows for which the starting point is January of that year. The horizontal line present on every plot indicates the 0.05 significance level. \\ 
\end{tabular}
  \end{center}
 \label{fig:rolling windows}
\end{figure}

\noindent
We test if the returns are elliptically symmetric in different time periods using a rolling window analysis. The window has a size of one year and it is rolled every month, \mbox{i.e.} we start with the window January 2000 - December 2000 and we test for elliptical symmetry. Then we shift the starting point by one month, that is we consider February 2000 - January 2001 and we test again for elliptical symmetry. We keep doing this until the last possible window. The following tests are used for every window: the test by Koltchinskii and Sakhanenko with \code{R = 100} bootstrap replicates, the MPQ test, Schott's test, the bootstrap test by Huffer and Park based on orthants with \code{c = 3} and  with the number of bootstrap replicates \code{R = 100}, the Pseudo-Gaussian test and the SkewOptimal test with the default values of the parameters. For every window we calculate the p-value. The results are presented in Figure ~\ref{rollingwindow}, where the horizontal line present on every plot indicates the 0.05 significance level. 

\noindent
Even though all these tests address the null hypothesis of elliptical symmetry, they have different powers for different alternative distributions and some tests may fail to detect certain departures from the null hypothesis. Certain tests are also by nature more conservative than others. We refer the reader to \cite{BGHL19} for a comparative simulation study that includes the majority of the tests available in this package.  This diversity in behavior presents nice opportunities. For instance, when all tests agree, we can be pretty sure about the nature of the analyzed data. One could also combine the six tests into a multiple testing setting by using a Bonferroni correction, though this is not what we are doing here. 

\noindent
The following general conclusions can be drawn from Figure ~\ref{rollingwindow}.
\begin{itemize}
\item In the past 20 years, the return data do not follow the elliptical distribution at least half of time. In other words, there are  many periods between 2000 and 2020 where the data exhibit some form of skewness or other type of symmetry, invalidating thus the mean-variance analysis.
\item The broader periods where the hypothesis of elliptical symmetry cannot be rejected are 2000-2004, 2005-2006, 2012-2013, 2015-2017 (for Schott's test only 2015-2016). In these periods, the tests may have only occasional rejections without a longer time period of rejections. \item In the period around the financial crisis in 2008, almost all tests reject the null hypothesis of ellipticity. This clearly shows that, in the periods of crisis, the assumption of elliptical symmetry is less likely to hold. 
\end{itemize}

\noindent
With the aim of guiding the reader through the functions that are available in the \pkg{ellipticalsymmetry} package, we now focus on the window January 2008 - December 2008. We start with the test by Koltchinskii and Sakhanenko.
\begin{verbatim}
> KoltchinskiiSakhanenko(data2008, R = 100)
\end{verbatim}

\begin{verbatim}
Test for elliptical symmetry by Koltchinskii and Sakhanenko

data:  data2008
statistic =  6.0884, p-value = 0.02
alternative hypothesis: the distribution is not elliptically symmetric
\end{verbatim}

\noindent
The \verb+KoltchinskiiSakhanenko+ output is simple and clear. It reports the value of the test statistic and p-value. For this particular data set the test statistic is equal to $6.0884$ and the p-value is $0.02$. Note that here we specify the number of bootstrap replicates to be \code{R = 100}. 

The MPQ test and Schott's test can be performed by running very simple commands: 

\begin{verbatim}
> MPQ(data2008) 
\end{verbatim}

\begin{verbatim}
Test for elliptical symmetry by Manzotti et al.

data:  data2008
statistic = 25.737, p-value = 0.04047
alternative hypothesis: the distribution is not elliptically symmetric
\end{verbatim}

\begin{verbatim}
> Schott(data2008)
\end{verbatim}

\begin{verbatim}
	Schott test for elliptical symmetry

data:  data2008
statistic = 24.925, p-value = 0.03532
alternative hypothesis: the distribution is not elliptically symmetric
\end{verbatim}

\noindent
Given the number of the input arguments, the function for the test by Huffer and Park deserves some further comments. The non-bootstrap version of the test can be performed by running the  command  
\begin{verbatim}
> HufferPark(data2008, c = 3)
\end{verbatim}

\begin{verbatim}
Test for elliptical symmetry by Huffer and Park

data:  data2008
statistic = 24.168, p-value = 0.109
alternative hypothesis: the distribution is not elliptically symmetric
\end{verbatim}

\noindent
By specifying \code{R} the bootstrap will be applied:
\begin{verbatim}
> HufferPark(data2008, c= 3, R = 100)
\end{verbatim}

\noindent
The p-value for the bootstrap version of the test is equal to $0.11$. Note that in both cases we used the default value for \code{sector}, that is \code{"orthants"}. 
\begin{verbatim}
	Test for elliptical symmetry by Huffer and Park

data:  data2008
statistic = 24.168, p-value = 0.11
alternative hypothesis: the distribution is not elliptically symmetric
\end{verbatim}

\noindent
If we want to change the type of sectors used to divide the space, we can do it by running the  command
\begin{verbatim}
HufferPark(data2008, c=3, R = 100, sector = "permutations")
\end{verbatim}
This version yields a p-value equal to $0.16$. 

\medskip
\noindent
Another very easy-to-use test is the Pseudo-Gaussian test:
\begin{verbatim}
> PseudoGaussian(data2008)
\end{verbatim}

\begin{verbatim}
	Pseudo-Gaussian test for elliptical symmetry

data:  data2008
statistic = 9.486, p-value = 0.02348
alternative hypothesis: the distribution is not elliptically symmetric
\end{verbatim}

\noindent
Eventually,  the following simple command will run the \code{SkewOptimal} test based on the radial density of the multivariate $t$ distribution with $4$ degrees of freedom  (note that the degrees of freedom could be readily changed by specifying the \code{param} argument).

\begin{verbatim}
> SkewOptimal(data2008)
\end{verbatim}

\begin{verbatim}
 SkewOptimal test for elliptical symmetry

data:  data2008
statistic = 12.209, p-value = 0.006701
alternative hypothesis: the distribution is not elliptically symmetric
\end{verbatim}

\noindent
The test based on the radial density of the multivariate logistic distribution can be performed by simply adding 
\code{f = "logistic"}:
\begin{verbatim}
> SkewOptimal(data2008, f = "logistic")
\end{verbatim}

\noindent
This version of the SkewOptimal test yields a p-value equal to $0.0003484$. Finally, if we want to run the test based on the radial density of the multivariate power-exponential distribution, we have to set \code{f} to \code{"powerExp"}. The kurtosis parameter equal to $0.5$ will be used unless specified otherwise. 
\begin{verbatim}
> SkewOptimal(data2008, f = "powerExp")
\end{verbatim}
The resulting p-value equals $0.002052$.
The kurtosis parameter can be changed by assigning a different value to \code{param}. For example,

\begin{verbatim}
SkewOptimal(data2008, f = "powerExp", param = 1.2)
\end{verbatim}

\noindent
We can conclude that the null hypothesis is rejected at the $5\%$ level by all tests except by Huffer and Park's tests. Luckily the tests available in the package mostly agree. In general, in situations of discordance between two (or more) tests, a practitioner may compare the essence of the tests as described in this paper and check if, perhaps, one test is more suitable for the data at hand than the other (e.g., if assumptions are not met). The freedom of choice among several tests for elliptical symmetry is an additional feature of our new package.

\section{Conclusion} \label{sec:summary}
In this paper, we have described several existing tests for elliptical symmetry and explained in details their \textbf{R} implementation in our new package \pkg{ellipticalsymmetry}. The implemented functions are simple to use, and we illustrate this via a real data analysis.  
The availability of  several tests for elliptical symmetry is clearly an appealing strength of our new package.


\section*{Acknowledgments}

Sla\dj ana Babi\' c was supported by a grant (165880) as a PhD Fellow of the Research Foundation-Flanders (FWO). Marko Palangeti\' c was supported by the Odysseus program of the Research Foundation-Flanders.

\bibliography{RJreferences}

\end{document}